# Weyl and Dirac Semimetals for Thermoelectric Applications

Saurabh Singh[1,*], Sarmistha Das[2,$], and Shiv Kumar[3]

[1)] Quantum Dynamics Unit, Okinawa Institute of Science and Technology (OIST) Graduate University, Onna, 904-0495 Okinawa, Japan
[2)] Department of Physics, Center for Advanced Nanoscience, University of California, San Diego, California 92093, USA
[3)] Institute of Microelectronics, Agency for Science, Technology, and Research (A*STAR), 138634, Singapore

Email: (*) saurabhsingh950@gmail.com, $ sdas@physics.ucsd.edu

Weyl and Dirac semimetals, characterized by their unique band structures with linear energy dispersion ($E$ vs $k$) near the Fermi level ($E_F$), have emerged as promising candidates for next-generation technology based on thermoelectric materials. Their exceptional electronic properties, notably high carrier mobility and substantial Berry curvature, offer the potential to surmount the limitations inherent in conventional thermoelectric materials. A comprehensive understanding of the fundamental physics underlying these materials is essential. This chapter mainly focused into the topological properties and distinctive electronic band structures of Weyl and Dirac semimetals, providing a theoretical framework for comprehending their thermoelectric transport properties such as Seebeck coefficients, electrical and thermal conductivity. The pivotal role of Berry curvature in enhancing Seebeck coefficients while reducing thermal conductivity is a key focus.

Experimental advancements in synthesizing single crystals and characterizing these materials have been significant. Recent development in material growth and characterization techniques have propelled research forward. The intricate relationship between material properties, such as carrier concentration, electronic bandgap, and crystal structure, and thermoelectric performance is explored. Realizing the potential of Weyl and Dirac semimetals for practical thermoelectric applications necessitates overcoming specific challenges. This chapter outlines strategies to optimize thermoelectric figures of merit ($ZT$) through band engineering, carrier doping, and nanostructuring. Moreover, the exploration of hybrid materials and heterostructures offers promising avenues for enhancing thermoelectric performance for renewable energy applications.

To fully harness the potential of these materials, continued research and development are imperative. This chapter concludes by providing an outlook on future research directions, encompassing the exploration of novel material systems, the development of advanced characterization techniques, and the integration of Weyl and Dirac semimetals into practical thermoelectric devices. By offering a comprehensive exploration of the fundamental physics, materials science, and device engineering aspects of Weyl and Dirac semimetals for thermoelectrics, this chapter aims to inspire further research and development in this dynamic field suitable to deal the modern technological challenge of thermal management.

# 1. Introduction

Energy is a fundamental requirement for any human society. The naturally available resources that fulfil the major energy requirements are coal, oil, and natural gas. These resources are limited in quantity, and it takes millions of years to form through natural processes. Modern society demands a huge consumption of energy due to the exponential increase in the number of electronic devices, motor vehicles, and the operation of small household appliances to large industrial equipment, which run day and night to support a sustainable society. The primary form of energy utilized is electricity, which is mostly generated from coal, oil, and natural gas, resulting in significant pollution and waste heat [1-4]. In the utilization, conversion, and consumption of electricity, up to 66% of the energy is wasted as heat, while only 33% is effectively utilized for its intended purpose [5,6].

According to the database, approximately 183,233 TWh of energy was consumed worldwide in 2023 [7]. Of this, approximately 77% was derived from non-renewable resources (coal: 45,565 TWh, oil: 54,565 TWh, natural gas: 40,102 TWh), and this consumption is expected to increase further in the coming years due to the growing global population [7]. The consumption of such a vast number of non-renewable resources produces toxic gases that are detrimental to human health and environmental well-being [8]. Another consequence of increasing pollution is adverse climate change, leading to a global average temperature increase of 1.5 degrees Celsius (as of 2024) compared to pre-industrial levels [9, 10]. As a result, severe effects on human health, reduced productivity, and environmental degradation are observed, manifested in the form of rising temperatures, sea-level rise, extreme weather events, damage to ecosystems, ocean acidification, etc [8]. These alarming situations necessitate the development and implementation of environmentally friendly, non-toxic, renewable, and sustainable energy sources [11-15].

An alternative solution of the above-mentioned problem is to explore renewable energy resources that can generate clean electricity from waste heat. In this direction, thermoelectric techniques, which allow the conversion of waste heat to electricity or vice versa [16-20], are discussed in more detail at the beginning of this chapter. Furthermore, this chapter discusses the criteria for identifying high-performance thermoelectric materials, their physical properties, and the potential of new material classes, including those based on Nernst and Ettingshausen effects and topological materials such as Weyl and Dirac semimetals. The aim of this chapter is to gain an overview of energy generation for modern technological applications.

# 2. Thermoelectric Effects in Topological Materials

## 2.1 Thermoelectric Effect:

The thermoelectric effect is a phenomenon that enables the conversion of waste heat into electricity, and vice versa, through the use of solid-state devices like thermoelectric modules, generators, and coolers [21, 22]. This conversion process, whether from heat to electricity or from electricity to cooling, relies on the Seebeck and Peltier effects, respectively [21-23]. Discovered in 1821 by German physicist Thomas Johann Seebeck, the Seebeck effect demonstrates that a potential difference arises across a metal object when a temperature gradient is applied [24]. In *n*-type metals, where free electrons are the dominant charge carriers, a temperature difference between the ends causes electrons with higher thermal energy to diffuse from the hotter end to the colder end. This electron movement establishes a thermo-emf voltage, $V_{th}$, at steady state across the metal, as shown schematically in Fig. 1a. For the *n*-type metal case, the direction of both heat flow (+Q to -Q) and the electric field are the same, but opposite to the direction of temperature gradient $\nabla T$. The ratio of $V_{th}$ to the temperature difference ($\Delta T$) is constant, known as the Seebeck coefficient or thermopower [22, 24, 25]. This coefficient, which varies depending on the material's electronic structure, carrier concentration, and temperature, indicates the material's ability to generate an electric potential difference per unit temperature difference. Characterizing these coefficients provides crucial information for understanding the mechanisms of charge transport within the material [26 - 28].

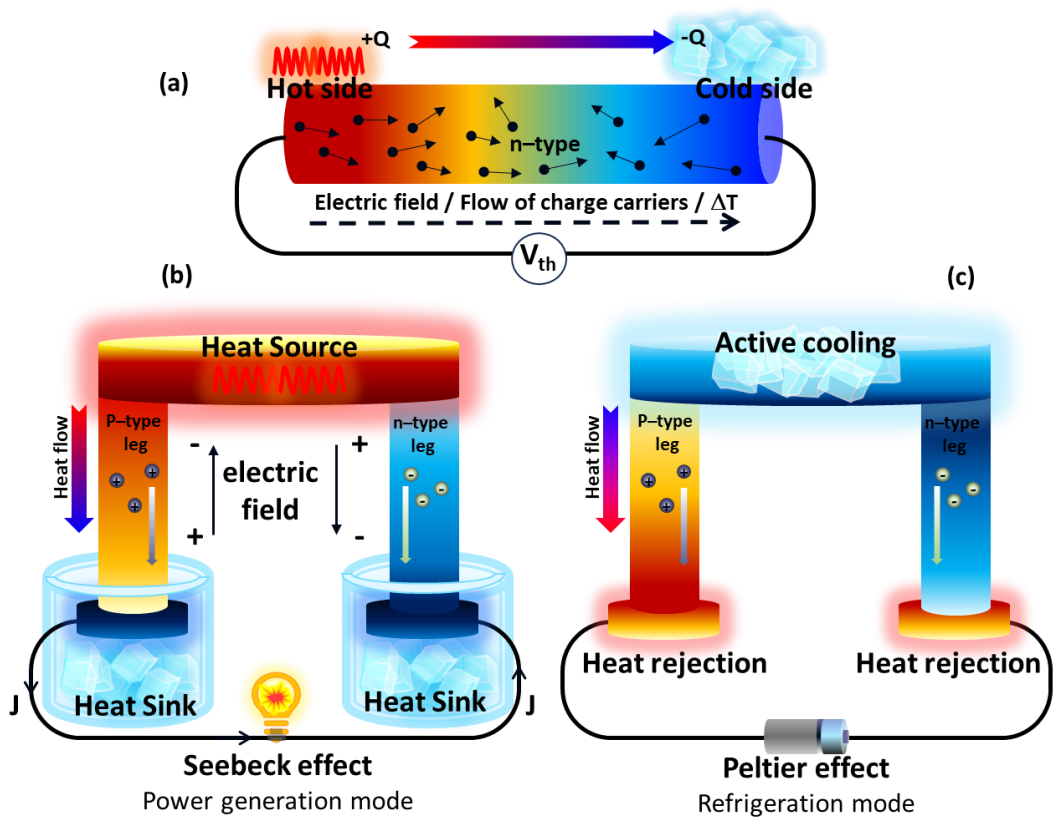

Fig. 1 Schematic of (a) the thermoelectric effect for an *n*-type metal, (b) the Seebeck effect in power generation mode, and (c) the Peltier effect in refrigeration mode for a pair of *p-n* leg devices.

Thermoelectric devices are fundamentally constructed using several number of *p*-type and *n*-type materials connected electrically in series while being subjected to a temperature gradient [29]. A typical thermoelectric device based on the Seebeck effect in the power generation mode is shown in Fig. 1b. When a sufficient temperature gradient is applied across the thermoelectric legs using a heat source and heat sink, electrons in n-type materials and holes in *p*-type materials move from the hot end to the cold end. This net flow of charge carriers results in the development of an electric field across the *p* and *n*-legs due to the buildup of thermo-emf voltage, leading to a net current flow (i.e., electric current density, *J*) that begins to flow through a load connected in series to the thermoelectric device. This current flowing through the closed circuit constitutes the output power generated by the thermoelectric device, which can be used to power electronic equipment [30, 31]. The generation of thermoelectric energy from waste heat has several advantages, such as being environmentally friendly, requiring low-cost maintenance, being quiet, compact, offering fast control, and being reliable for both temperature sensing and control [32].

In contrast to the Seebeck effect, the Peltier effect demonstrates the ability of thermoelectric devices to actively control heat flow. When connected in series with an external power supply, a thermoelectric device can absorb or liberate heat at the junction of the *p* and *n*-legs. The amount of heat absorbed or liberated, based on the Peltier effect (discovered by French physicist Jean Charles Athanase Peltier) [33], depends on the amount of current flowing through the device and the direction of that current flow.

The schematic of the Peltier effect is shown in Fig. 1c, where a battery is used to supply the current in the thermoelectric device, which acts as a driving force for the motion of charge carriers through *p-n* junctions, resulting in heat generation and heat absorption at the junction [34]. Therefore, thermoelectric devices can also be used for solid-state refrigeration. Solid-state cooling/heating thermoelectric devices are very useful as they are easy to control heating and cooling, offering fast response, which is essential in medical devices and consumer electronics [35-37].

## 2.2 Thermoelectric materials

To determine the suitability of a material for thermoelectric applications, it is crucial to assess its thermoelectric conversion efficiency ($\eta$), which is primarily quantified based on the material's physical properties using a dimensionless quantity called as *figure of merit*, *ZT*. The mathematical expression for thermoelectric efficiency [14, 16, 34] in terms of *ZT* is described by the following equation:

$$\eta = \frac{T_H - T_C}{T_H} \frac{\sqrt{1+ZT_{avg}}-1}{\sqrt{1+ZT_{avg}}+T_C/T_H} \quad\text{------------------------------------------------(1)}$$

where, $T_H$ is the hot junction temperature, $T_C$ represents the cold junction temperature, and $ZT_{avg}$ denotes the average *figure of merit*. The initial term on the right-hand side, i.e. $\frac{T_H - T_C}{T_H}$, of equation (1) corresponds to the Carnot efficiency ($\eta_{Carnot}$), which is mainly depends upon the temperature differential across the thermoelectric generator (TEG). The subsequent term represents the reduction factor, which is a function of the material's *figure of merit*, specifically determined by the average *ZT* ($ZT_{avg}$). Using the high-quality modern tools of temperature sensing technology, temperature at the hot ($T_H$) and cold ends ($T_C$) can be measured with millikelvin accuracy. Therefore, an accurate estimation of the thermoelectric generator's (TEG's) conversion efficiency necessitates meticulous evaluation of the average *figure of merit* for both *n*-type and *p*-type materials.

The *figure-of-merit* (*ZT*) is a critical material parameter that significantly influences thermoelectric efficiency. The *ZT* value for a specific material can be determined by measuring its Seebeck coefficient (*S*), electrical resistivity ($\rho$), and thermal conductivity ($\kappa$). Mathematically, the expression for *ZT* [14, 16, 34] at absolute temperature (*T*) can be represented as:

$$ZT = \left(\frac{S^2}{\rho \cdot \kappa}\right) T \text{-----------------------------------------------------------(2)}$$

Here, $T$ represents the average sample temperature, which is calculated as the mean of the hot end and cold end temperatures of the sample used for Seebeck coefficient measurements.

It is clearly evident from equation (2) that for a high $ZT$, the material should exhibit a large Seebeck coefficient ($S$), low electrical resistivity (or high electrical conductivity, $\sigma = 1/\rho$), and low thermal conductivity ($\kappa$). In metals, semimetals, and semiconductors, charge carriers are the primary contributors to charge transport properties. However, in the total thermal conductivity, both charge carriers and the lattice contribute. The total thermal conductivity can be expressed as $\kappa_e + \kappa_l$, where $\kappa_e$ and $\kappa_l$ represents the electronic and lattice thermal conductivities, respectively. The electrical conductivity and electronic thermal conductivity are interrelated through the Wiedemann–Franz law, i.e., $\kappa_e = L \cdot \sigma \cdot T$, where $L$ is the constant known as the Lorentz number. For most metals in the degenerate limit, $L$ typically has values of ~2.44 × $10^{-8}$ $V^2K^{-2}$, as they exhibit small Seebeck coefficients (~1 – 30 $\mu V/K$) [25, 34]. However, for the case of non-degenerate semiconductors, a typical range of Seebeck coefficients is ~50 to 500 $\mu V/K$ [28]. Therefore, the values of the Lorentz number need to be evaluated using the appropriate formula $L = 1.5 + exp\left[-\frac{|S|}{116}\right]$ as described by Kim *et al.* [38].

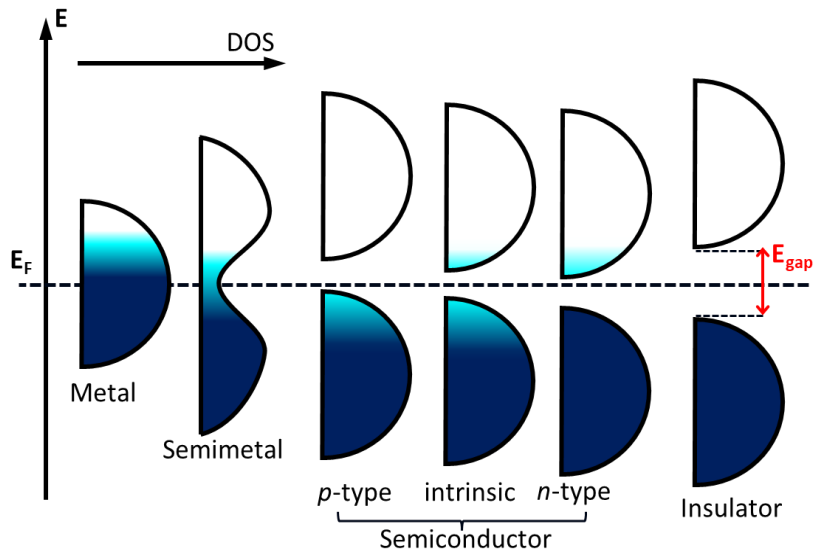

Fig. 2 Schematic representation of the electronic density of states (DOS) for metal, semimetal, intrinsic semiconductor, *p*-type semiconductor, *n*-type semiconductor, and insulator.

In the search for high-performance thermoelectric materials, it is essential to simultaneously optimize three key physical properties i.e. Seebeck coefficient ($S$), electrical conductivity ($\sigma$), and thermal conductivity ($\kappa$), to achieve a high $ZT$ value. However, this optimization is challenging because these physical parameters are interdependent, often through a common physical quantity such as carrier concentration ($n$) [25]. From an electronic structure perspective, an optimized carrier concentration (in heavily doped semiconductors) with high mobility can be a promising approach, particularly when the material's inherent lattice thermal conductivity is lower than 1 $W \cdot m^{-1} \cdot K^{-1}$ [25, 27].

To better understand and select suitable candidates, it is crucial to consider the electronic density of states (DOS) and band dispersion ($\varepsilon$ vs. $k$) of the material. The electronic density of states for metals, semimetals, semiconductors, and insulators is shown in Fig. 2. For the metal case, highly dispersive bands are found to intersect the Fermi level ($E_F$), resulting in a high density of states (DOS) at $E_F$ and consequently high electrical conductivity. Whereas for the semimetal case, few energy bands intersect the Fermi level, resulting in a finite density of states, but much lower compared to that in the metal case [21, 25, 27].

Unlike metals and semimetals, there are other classes of materials that exhibit no density of states in the vicinity of the Fermi level, instead having a finite energy band gap formed between the top of the valence band and the conduction band. Materials with narrow band gaps ($E_{gap}$ ~0.2 – 1.5 eV) are generally classified as intrinsic semiconductors, where the Fermi level is located at the center of the band gap. The electronic structure of these materials can be further modified by introducing electron ($n$-type) or hole ($p$-type) dopants, which shifts the Fermi level position towards the conduction band or valence band, respectively. Materials with larger band gaps, typically in the range of 2.0 eV to 5.0 eV, are generally classified as insulators. These materials exhibit very low carrier concentrations, and it is quite challenging to tune their electronic structures except under extreme conditions such as high pressure or alloying.

From the perspective of thermoelectric applications, metals are primarily used in thermocouples to sense temperature by utilizing the Seebeck effect between two dissimilar metals (one with a positive Seebeck coefficient and the other with a negative Seebeck coefficient) joined at one end [39]. This is primarily due to their ductility, long durability, low cost and high thermal conductivity (~100 – 500 W m$^{-1}$K$^{-1}$), which make them well-suited for sensing the temperature of a localized point with a very fast response and precise control (up to few millikelvin) [25, 40 – 42]. However, due to their very high carrier concentrations ($10^{22}$ – $10^{23}$ cm$^{-3}$), metals are not ideal for use in thermoelectric generators or coolers, as their $ZT$ values are very low, resulting in insignificant conversion efficiency [25]. Considering the longitudinal Seebeck effect (shown in Fig.1a), both metals and semimetals are generally not considered suitable for energy conversion devices, unless a low magnitude of thermal conductivity is achieved by making a nanowires or atomic layer thin films [22, 25, 26, 28]. However, the recent discovery of new classes of materials such as Weyl and Dirac semimetals, which exhibit high power factors based on the Nernst effect (to be discussed later), offers new possibilities. Such high-power factors can be obtained by applying an external magnetic field in addition to the temperature gradient across these materials.

The most promising materials showing high $ZT$ and have potential for making the thermoelectric generator applications belong to the family of semiconductors. From a fundamental perspective, the magnitude of the Seebeck coefficient ($S$) is directly related to the effective mass of charge carriers. This relationship, within the framework of the free electron approximation, can be expressed as: [14, 16, 25, 27, 28]

$$S = \left(\frac{8\pi^2 k_B^2}{3eh^2}\right) m^* T \left(\frac{\pi}{3n}\right)^{2/3} \text{-----------------------------(3)}$$

where, $h$ is Planck's constant, $k_B$ is Boltzmann constant, T is the absolute temperature, which is an independent constant factor for any material. In equation (3), the material-dependent electronic parameters are $m^*$ and $n$, representing the effective mass of charge carriers and carrier density, respectively. Band engineering approaches provide a means to tune $m^*$ and $n$, enabling the optimization of the Seebeck coefficients of degenerate semiconductors through precise control of these two physical quantities with suitable doping. The effective mass [25],

$m^* = \left(\frac{\hbar^2}{\frac{d^2E}{dk^2}}\right)$, obtained from the electronic band structure, provides important information for understanding the Seebeck coefficient behavior and the type of dominant charge carrier. The large magnitude of the effective mass of electrons (*n*-type) or holes (*p*-type) typically determines the sign of the Seebeck coefficient of a material, with negative values for *n*-type materials and positive values for *p*-type materials [43 – 47].

Materials with better performance can be screened by performing electronic structure calculations using density functional theory-based tools. The electronic transport coefficients ($\sigma(T)$, $S(T)$, and $\kappa_{el}(T)$), based on the linear response theory implemented within the Kubo-Greenwood formula and Boltzmann transport theory [48 – 51], can be written as:

$$\sigma(T) = \int_{-\infty}^{\infty} \sigma(\varepsilon, T)\left(-\frac{\partial f_{FD}(\varepsilon, T)}{\partial \varepsilon}\right) d\varepsilon \text{------------------------(4)}$$

$$S(T) = -\frac{1}{|e|T} \frac{\int_{-\infty}^{\infty}(\varepsilon-\mu)\sigma(\varepsilon,T)\left(-\frac{\partial f_{FD}(\varepsilon,T)}{\partial \varepsilon}\right)d\varepsilon}{\int_{-\infty}^{\infty}\sigma(\varepsilon,T)\left(-\frac{\partial f_{FD}(\varepsilon,T)}{\partial \varepsilon}\right)d\varepsilon} \text{------------------------(5)}$$

$$\kappa_{el}(T) = \frac{1}{e^2 T}\int_{-\infty}^{\infty}(\varepsilon-\mu)^2 \sigma(\varepsilon, T)\left(-\frac{\partial f_{FD}(\varepsilon, T)}{\partial \varepsilon}\right)d\varepsilon - \frac{1}{e^2 T}\frac{\left\{\int_{-\infty}^{\infty}(\varepsilon-\mu)\sigma(\varepsilon,T)\left(-\frac{\partial f_{FD}(\varepsilon,T)}{\partial \varepsilon}\right)d\varepsilon\right\}^2}{\int_{-\infty}^{\infty}\sigma(\varepsilon,T)\left(-\frac{\partial f_{FD}(\varepsilon,T)}{\partial \varepsilon}\right)d\varepsilon} \text{---(6)}$$

where, $\sigma(\varepsilon, T) = \frac{|e|^2}{3} N(\varepsilon) v^2 \tau(\varepsilon, T)$ is the spectral conductivity which depends on the density of states $N(\varepsilon)$, group velocity ($v$), and relaxation time $\tau(\varepsilon, T)$. The energy and temperature dependent parameter in equation (4-6) is $f_{FD}(\varepsilon, T) = \frac{1}{e^{(\varepsilon-\mu)/k_B T} + 1}$ known as Fermi-Dirac distribution function, which depends upon the chemical potential ($\mu$), provide the probability of electron occupancy in energy states above and below the Fermi energy. Equation (4-6) suggests that, within the strong scattering limit, materials with a large variation in the density of states near the band edge within an energy range of 10 $k_B T$ can exhibit both high electrical conductivity and high Seebeck coefficients [43- 48]. Therefore, a constructive modification of the electronic structure, such as creating impurity states near the band edges, is an essential criterion for achieving a large power factor ($S^2\sigma$) [48 – 56]. This has been experimentally realized in several materials, including Si-Ge, silver chalcogenides, and tellurides [57 – 62]. In addition to the power factor, an important parameter which can be independently controllable is the lattice thermal conductivity for achieving the high *ZT*.

Lattice thermal conductivity [25] can be written as:

$$\kappa_l = \frac{1}{3} C_v V_g\, l = \frac{1}{3} C_v V_g^{\,2}\, \tau \text{------------------------(6)}$$

where, $C_v$, $V_g$, $l$ and $\tau$ are the specific heat at constant volume, phonon group velocity, phonon mean free path, and phonon relaxation time, respectively. Equation (6) suggests that low lattice thermal conductivity can be obtained in materials with low specific heat and a low mean free path resulting from strong phonon scattering caused by lattice defects, heavy elements, and grain boundaries, etc. Several materials possess ultra-low thermal conductivity due to the complex crystal structure and strong anharmonic scattering effects [25, 53]. For such materials, modifying the electronic structure can help to achieve a large *ZT* value [55, 56]. On the other hand, materials with high thermal conductivity and a large power factor can be engineered into nanostructures through techniques like mechanical ball milling and chemical synthesis to suppress thermal conductivity and achieve a high *ZT* value [58, 60]. Therefore, optimizing both electronic and lattice transport properties is a crucial criterion for achieving a high figure of merit (*ZT* > 1) and high conversion efficiency ($\eta$ > 15%) [27, 28, 41], which are essential for developing effective and affordable thermoelectric devices for waste heat management.

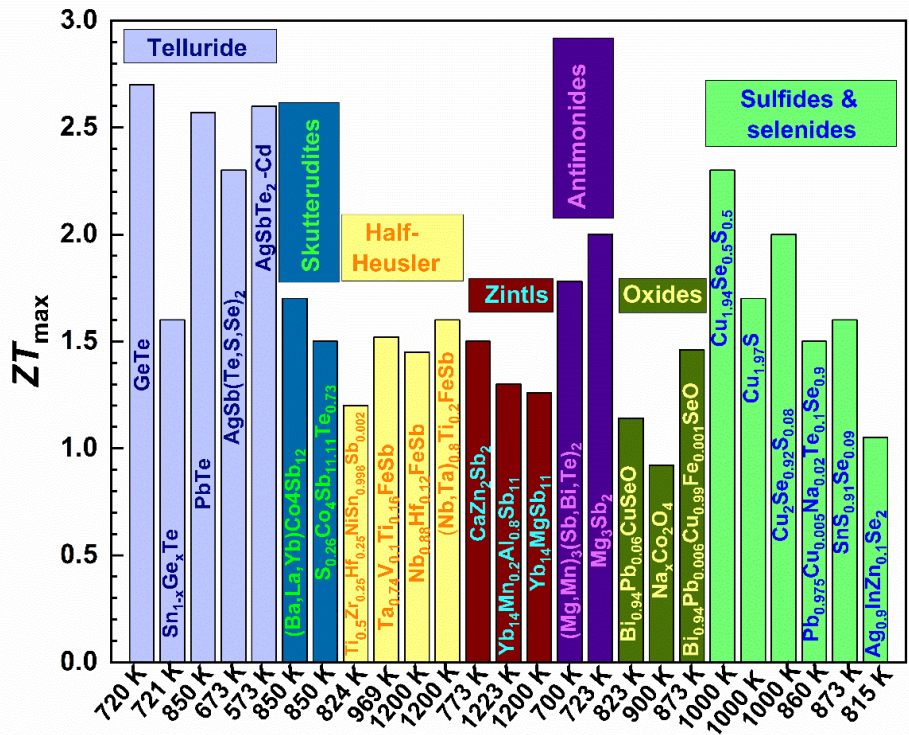

Fig. 3 Maximum reported *figure of merit* (*ZT*) for *state-of-the-art* thermoelectric materials.

Although the Seebeck effect was discovered over 200 years ago [24], significant research in thermoelectricity did not flourish due to low thermoelectric properties and conversion efficiencies. With the development of new and innovative materials synthesis techniques, state-of-the-art characterization tools, and high-performance computational tools for electronic and phonon structure calculations, a dedicated effort has been undertaken by the thermoelectric research community in the past three decades to discover new and high-performance thermoelectric materials with high *ZT* values [27, 28].

In this direction, several different classes of materials have been investigated, including silicides, oxides, sulfides, selenides, actinides, lanthanides, borides, carbides, nitrides, $FeGa_3$-type materials, antimonides, Zintl phases, clathrates, half-Heuslers, tellurides, and skutterudites [27, 28, 29, 52, 63]. Figure 3 shows the maximum reported figure of merit (*ZTmax*) for selected thermoelectric materials that have been found to be suitable candidates for thermoelectric applications in the low, mid, and high-temperature ranges. The composition details and the temperature at which maximum *ZT* is obtained for the materials are shown in Table 1. Among the possible candidates, $AgSbTe_2$, $PbTe$, $GeTe$, $Cu_2Se$, and $Mg_3Sb_2$-based materials show a maximum $ZT \geq 2.0$, which is a considerable magnitude for achieving reasonable conversion efficiency in thermoelectric applications.

Table 1. $ZT$ performance of *state-of-the-art* thermoelectric materials, Ref. [64 – 90].

| Sr. No. | Materials | $ZT$ | Temperature (K) | Ref. |
|---|---|---|---|---|
| 1 | GeTe | 2.7 | 720 | [64] |
| 2 | $Sn_{1-x}Ge_xTe$ | 1.6 | 721 | [65] |
| 3 | PbTe | 2.57 | 850 | [66] |
| 4 | $AgSb(Te,S,Se)_2$ | 2.3 | 673 | [67] |
| 5 | $AgSbTe_2$-Cd | 2.6 | 573 | [68] |
| 6 | $(Ba,La,Yb)Co_4Sb_{12}$ | 1.7 | 850 | [69] |
| 7 | $S_{0.26}Co_4Sb_{11.11}Te_{0.73}Te_{0.73}$ | 1.5 | 850 | [70, 71, 72] |
| 8 | $Ti_{0.5}Zr_{0.25}Hf_{0.25}NiSn_{0.998}Sb_{0.002}$ | 1.2 | 824 | [73] |
| 9 | $Ta_{0.74}V_{0.1}Ti_{0.16}FeSb$ | 1.52 | 969 | [74] |
| 10 | $Nb_{0.88}Hf_{0.12}FeSb$ | 1.45 | 1200 | [75] |
| 11 | $(Nb_{0.6}Ta_{0.4})_{0.8}Ti_{0.2}FeSb$ | 1.60 | 1200 | [76] |
| 12 | $CaZn_2Sb_2$ | 1.5 | 773 | [77] |
| 13 | $Yb_{14}Mn_{0.2}Al_{0.8}Sb_{11}$ | 1.3 | 1223 | [78] |
| 14 | $Yb_{14}MgSb_{11}$ | 1.26 | 1200 | [79] |
| 15 | $(Mg,Mn)_3(Sb,Bi,Te)_2$ | 1.78 | 700 | [80] |
| 16 | $Mg_3Sb_2$ | 2.0 | 723 | [81] |
| 17 | $Bi_{0.94}Pb_{0.06}CuSeO$ | 1.14 | 823 | [82] |
| 18 | $Na_xCo_2O_4$ | 0.92 | 900 | [83] |
| 19 | $Bi_{0.94}Pb_{0.006}Cu_{0.99}Fe_{0.001}SeO$ | 1.46 | 873 | [84] |
| 20 | $Cu_{1.94}Se_{0.5}S_{0.5}$ | 2.3 | 1000 | [85] |
| 21 | $Cu_{1.97}S$ | 1.7 | 1000 | [86] |
| 22 | $Cu_2Se_{0.92}S_{0.08}$ | 2.0 | 1000 | [87] |
| 23 | $Pb_{0.975}Cu_{0.005}Na_{0.02}Te_{0.1}Se_{0.9}$ | 1.5 | 860 | [88] |
| 24 | $SnS_{0.91}Se_{0.09}$ | 1.6 | 873 | [89] |
| 25 | $Ag_{0.9}InZn_{0.1}Se_2$ | 1.05 | 815 | [90] |

For the mid-temperature range, tellurides, skutterudites, and antimonides are the best choices, whereas for the high-temperature range, half-Heuslers, Zintl phases, and oxides are considered more suitable due to their high performance, chemical, and thermal stability. Around room temperature, one of the key applications is to provide input power supply using wearable thermoelectric generators made of ductile inorganic materials [91-93]. In this direction, high-performance and flexible silver chalcogenide ($Ag_2S$) based materials and devices have been discovered in recent years [61, 62, 94 – 96].

Although chalcogenides have shown a very prominent role in thermoelectric performance across low, mid-, and high-temperature ranges, the toxicity of constituents such as tellurium and its scarcity pose significant challenges for large-scale commercialization. In comparison to tellurium-based thermoelectric alloys [64 – 68], potential alternatives for high-temperature applications include Si-Ge and Heusler alloys [57, 58, 60, 73, 74, 76], with $ZT$ values exceeding 1.5. For room-temperature applications, $Ag_2S$-based flexible thermoelectrics could be an excellent choice for providing input power to large-scale wearable electronic sensors and devices [61, 62, 94 – 96].

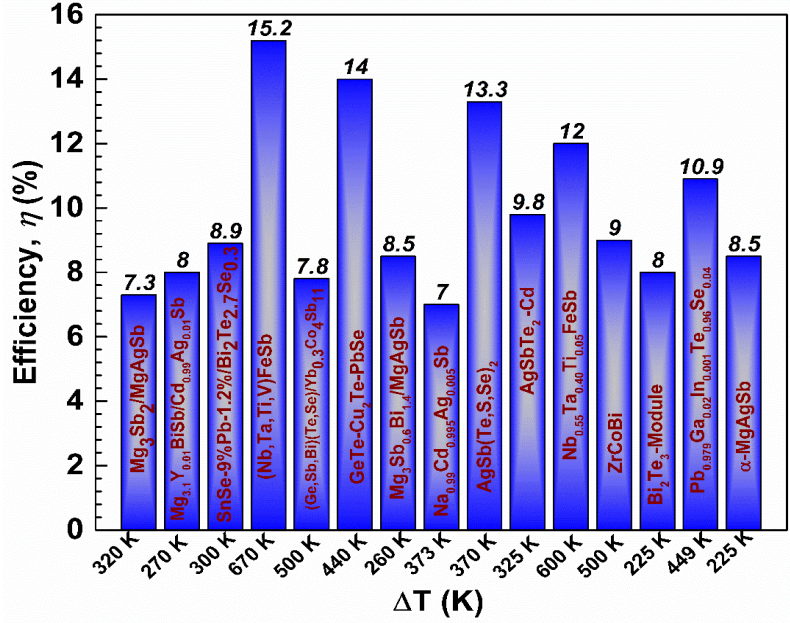

Fig. 4 Comparison of Maximum Efficiency for Various Thermoelectric Modules

Table 2. Conversion efficiency, $\eta$, reported for *state-of-the-art* thermoelectric materials

| Sr, No. | Material | TE Efficiency, $\eta$ (%) | Temperature Difference $\Delta T$ (K) | Ref. |
|---|---|---|---|---|
| 1 | $Mg_{3.1}Sb_2$/MgAgSb | 7.3 | 320 | [97] |
| 2 | $Mg_{3.1}Y_{0.01}BiSb$/ $Cd_{0.99}Ag_{0.01}Sb$ | 8 | 270 | [98] |
| 3 | SnSe-9%Pb-1.2%/$Bi_2Te_{2.7}Se_{0.3}$ | 8.9 | 300 | [99] |
| 4 | (Nb,Ta,Ti,V)FeSb | 15.2 | 670 | [100] |
| 5 | $Ge_{0.92}Sb_{0.04}Bi_{0.04}Te_{0.95}Se_{0.05}$/ $Yb_{0.03}Co_4Sb_{11}$ | 7.8 | 500 | [101] |
| 6 | GeTe-$Cu_2$Te-PbSe | 14 | 440 | [102] |
| 7 | $Mg_3Sb_{0.6}Bi_{1.4}$/MgAgSb | 8.5 | 260 | [103] |
| 8 | $Na_{0.99}Cd_{0.995}Ag_{0.005}Sb$ | 7 | 373 | [104] |
| 9 | AgSb(Te,S,Se)$_2$ | 13.3 | 370 | [67] |
| 10 | AgSbTe$_2$ – Cd | 8 | 325 | [68] |
| 11 | $Nb_{0.55}Ta_{0.4}Ti_{0.05}FeSb$ | 12 | 600 | [105] |
| 12 | ZrCoBi | 9 | 500 | [106] |
| 13 | $Bi_2Te_3$-Module | 8 | 225 | [107] |
| 14 | $Pb_{0.979}Ga_{0.002}In_{0.001}Te_{0.96}Se_{0.04}$ | 10.9 | 449 | [108] |
| 15 | $\alpha$-MgAgSb | 8.5 | 225 | [109] |

The thermoelectric efficiency of single-leg, unicouple, and multileg modules is shown in Fig. 4. From the comparative plot, it is clearly evident that laboratory-based results have reported maximum efficiencies up to 15.2% for a temperature gradient of approximately 670 K applied across a module fabricated using (Nb, Ta, Ti, V) FeSb-based alloys [100]. Several chalcogenide- and antimonide-based modules with efficiencies in the range of 7-14% have also been reported [97 – 109], offering competitive alternatives to commercially available $Bi_2Te_3$-based modules, which have efficiencies limited to the 5-8% range [110, 111]. Table 2 summarizes the maximum thermoelectric efficiency obtained for different thermoelectric modules at the applied temperature gradient across the device.

$Bi_2Te_3$-based commercial devices are useful for near-room-temperature applications, however, toxic and scarcity tellurium prevent its use at large scale both from economical and safety point of view, especially the place of use where human involvement is in direct contact to the device. However, their performance starts to degrade when the temperature at the hot end of the device exceeds 200 degrees Celsius due to thermal and chemical stability issues. Therefore, it is highly desirable to search for high-performance thermoelectric materials that exhibit a high figure of merit across a wide temperature range including cryogenic temperature, near room temperature, and high temperature. In this direction, it is expected that several new materials, including oxides, Si-Ge alloys, heavy fermions, and Heusler alloys, will be discovered by utilizing the constructive approach in the coming future for the development of clean, non-toxic, cheap, and environmentally friendly thermoelectric modules for thermoelectric generation and cooling applications based on solid-state devices [112, 113, 114].

## 2.3 Topological Materials

A collection of materials in which the band structure of a typical insulator is inverted due to the spin-orbit coupling effect is classified as a topological insulator [115, 116]. This can be understood in a simple way as follows: a topological insulator possesses a negative band gap, as shown in Fig. 5a [117]. In other words, metallic surface states arise due to the topology of the inverted bulk band structure [118, 119]. Therefore, depending on the topological nature of the band structure, a bulk insulator can be further classified into different types of insulators, each exhibiting several exotic physical properties [120 -124]. Topological insulators are a subset of quantum materials exhibits unique properties arising primarily from topological properties. Topological insulators typically exhibit gapless surface states in the vicinity of the bulk energy band gap. Several materials ($Bi_2Se_3$ and HgTe) have been characterized as topological insulators, exhibiting band dispersions resembling a Dirac cone [125 – 132].

Similar to topological insulators, topological surface states have also been observed on the surfaces of several semimetals, which are further classified as Weyl semimetals. For a Weyl semimetal to exist, either time-reversal symmetry or inversion symmetry must be broken. In a material where both time-reversal symmetry and inversion symmetry coexist, the material possesses a pair of degenerate Weyl points, resulting in a new phase of topological semimetal known as the Dirac semimetal phase (shown in Fig. 5b) [117]. Furthermore, Weyl semimetals (WSMs) are classified into two categories, type I and type II, based on the energetics of the Fermi surface at the Weyl points (shown in Fig. 5c and 5d). In type-I WSMs, the Fermi surface intersects the Weyl points. In type-II WSMs, the Weyl point behaves like a touching point due to the strong tilting of the Weyl cone. This touching point acts as a common point between electron and hole pockets in the Fermi surface [133- 137].

From a materials perspective, initial theoretical calculations suggested the possibility of WSM in several materials, including $Hg_{1-x-y}Cd_xMn_yTe$, $Y_2Ir_2O_7$, and $HgCr_2Se_4$ [138-140]. However, these topological features were not observed experimentally. In the search for WSMs, several promising materials (NbP, TaP, TaAs, and NbAs) were found theoretically and further validated experimentally to exhibit Fermi arc features using ARPES (Angle-resolved photoemission spectroscopy), validating earlier theoretical predictions [141-145]. In these materials, the realization of WSM features from ARPES data

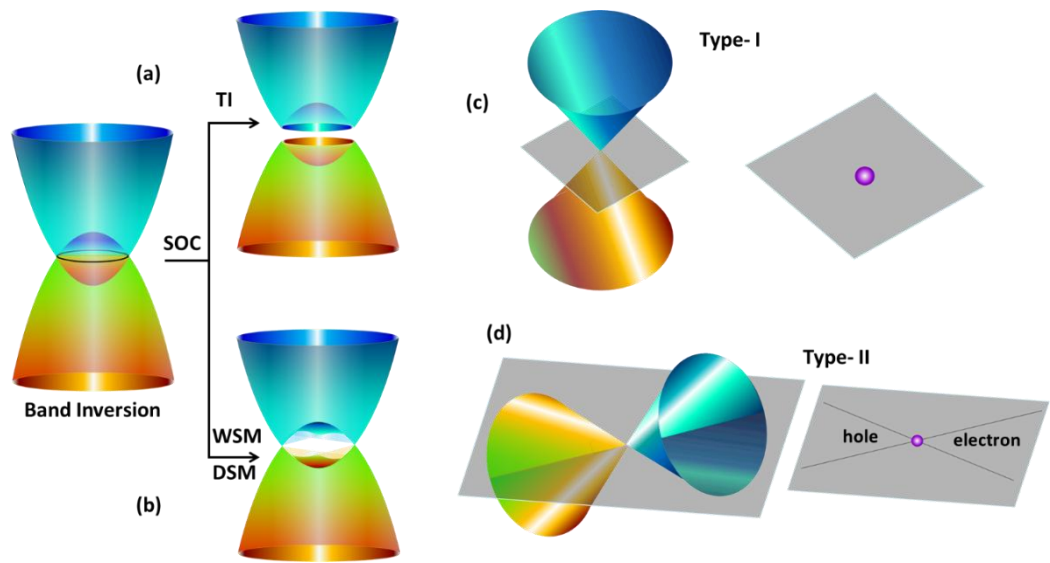

Fig. 5 (a) Topological Insulator with an open band gap after band inversion and spin orbit coupling effect, (b) Weyl semimetal and Dirac semimetals (c) Type- I WSM and (d) Type – II WSM.

helps to understand the magnetotransport properties, of these parent compound as well intermediate compositions, governed by the breaking of crystal inversion symmetry while preserving time-reversal symmetry. Furthermore, several Dirac semimetals ($Cd_3As_2$, $Na_3Bi$, $β$-CuI, BaAuSb) with promising electronic, optical, and magnetic properties have also been discovered through the combined use of experimental and theoretical tools [146- 151].

Recently, WSMs have been extensively investigated because of the giant response of the anomalous Nernst effect and Ettingshausen effect. The experimental observation of these effects, realized in type-II semimetals like $WTe_2$, opens up new techniques for harvesting waste heat to generate useful electricity. Utilizing the topological surface states and unique band structure features, WSMs provide control over the voltage generation, which builds up perpendicular to both the applied magnetic field and temperature gradient. Devices based on the Nernst effect are therefore very useful for detecting magnetic fields with high accuracy, as well as for potentially converting waste heat to electricity with a simple and low-cost design.

## 2.4 Nernst and Ettingshausen Effect

The Nernst effect was discovered in the early twentieth century by Walther Nernst [152]. This effect has since been utilized to understand the fundamental physics of topological materials and explore applications in waste heat recovery, similar to the Seebeck effect. The schematic of the Nernst effect is shown in Fig. 6a, where an external magnetic field ($B$) applied perpendicular to the temperature gradient ($\nabla T$) results in the generation of a transverse voltage perpendicular to both $B$ and $\nabla T$. This effect is analogous to the Hall effect, with the key difference being that in the Hall effect, an external current is applied to drive the charge carriers, while in the Nernst effect, the Seebeck electric field generated due to the temperature gradient serves a similar purpose. Here, it is important to note that in the longitudinal Seebeck effect (shown in Fig. 1a), the magnitude of the Seebeck coefficient is determined by the generated thermoelectric voltage caused by $\nabla T$. Therefore, a significant temperature gradient and a large sample size (for maintaining the temperature gradient) are essential for accurately probing the Seebeck coefficient. However, in the case of the Nernst effect, any temperature gradient is sufficient to probe the Nernst signal, as the strength of the measured voltage is primarily determined by the magnitude of the applied magnetic field. Therefore, there are no significant restrictions on sample dimensions or critical temperature gradients during the measurement. Such advantages are also beneficial for designing the device with small size [152].

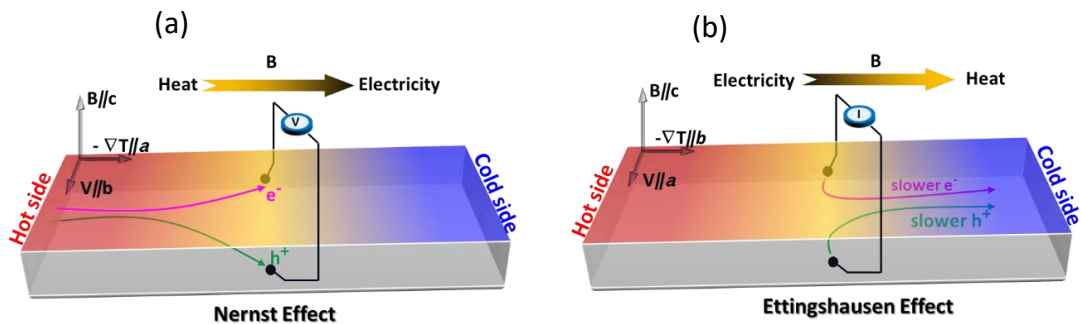

Fig 6 (a) Nernst Effect, (b) Ettingshausen effect

For a semimetal case, with two types of compensated charge carriers (electrons and holes), the expression for the Nernst coefficient, $N = (S_{yx}/B)$, can be written as

$$N = \frac{(N_e \sigma_e + N_h \sigma_h) + (\sigma_e + \sigma_h) + (N_e \sigma_e \mu_h - N_h \sigma_h \mu_e)(\sigma_e \mu_h - \sigma_h \mu_e)B^2 + \sigma_e \sigma_h (\mu_h + \mu_e)(\alpha_h - \alpha_e)}{(\sigma_e + \sigma_h)^2 + (\sigma_e \mu_h - \sigma_h \mu_e)^2 B^2} \quad \ldots\ldots\ldots 7)$$

where $\sigma_e$ and $\sigma_h$ are the electrical conductivities, $N_e$ and $N_h$ are the Nernst coefficients, $\alpha_e$ and $\alpha_h$ are the Seebeck coefficients, $\mu_e$ and $\mu_h$ are the carrier mobilities of electrons and holes, respectively; $B$ is the magnetic field; and $S_{yx}$ is the transverse Seebeck coefficient [154]. In the case where electrons and holes mobilities are comparable and have the similar carrier concentrations, such as $WTe_2$ system, the $B^2$ terms can be neglected for the expression given in equation (7).

The schematic of the cooling mechanism based on the Ettingshausen effect is shown in Fig. 6b. In this effect, a magnetic field ($B$) applied perpendicular to the current flow ($I$) results in the accumulation of charge carriers at one end (perpendicular to both $B$ and $I$), causing strong scattering of charge carriers and generating heat at this end (known as the hot end) due to an increased probability of carrier scattering. Therefore, the Ettingshausen effect can be utilized to convert electrical energy into heat in the perpendicular direction, leaving the other end at a lower temperature (known as the cold end) [155. 156]. Although Ettingshausen cooling devices were fabricated in the 1960s, their practical applications were limited due to low performance and the need for very strong magnetic fields for their operation [157, 158]. Taking advantage of new techniques, in recent years, several Weyl and Dirac semimetals have been explored, exhibiting high performance of Nernst Power factor, at low magnetic fields, which suggests their potential for large-scale energy generation and cooling applications.

Table 3: Measured Nernst Seebeck Coefficients ($S_{yx}$), Nernst Power Factor (*PF*), Applied Magnetic Field (*B*), and Measurement Temperature (*T*) for Selected Weyl Semimetals.

| Sr No. | Material | $S_{yx}$ (μV/K) | PF ($S_{yx}^2 \cdot \rho_{yy}$) (μW cm$^{-1}$ K$^{-2}$) | Magnetic Field (Tesla) | Temperature (K) | Ref. |
| --- | --- | --- | --- | --- | --- | --- |
| 1 | WTe$_2$ | 7000 | 30000 | 9 | 11.3 | [159] |
| 2 | NbSb$_2$ | 616 | 3800 | 9 | 21 | [160] |
| 3 | Mg$_2$Pb | 200 | 400 | 7 | 30 | [161] |
| 4 | PtSn$_4$ | 45 | 90 | 9 | 10.3 | [162] |
| 5 | Cd$_3$As$_2$ | 100 | 50 | 3 | 250 | [163] |
| 6 | ZrTe$_5$ | 1900 | 35 | 13 | 100 | [164] |
| 7 | NbP | 800 | 200 | 9 | 100 | [165] |
| 8 | Mg$_3$Bi$_2$ | 100 | 21.82 | 6 | 13.5 | [166] |
| 9 | NbAs$_2$ | 150 | 640 | 2 | 35 | [167] |
| 10 | NbAs$_2$ | 400 | 850 | 5 | 35 | [167] |

Table 3 provides a list of Weyl semimetals exhibiting the Nernst-Ettingshausen effect [159 – 167]. The best power factor is obtained for WTe$_2$ single crystals at 11.3 K and 9 tesla. The plots of Nernst power factor ($S_{yx}^2 \cdot \rho_{yy}$) as a function of Nernst Seebeck coefficient and applied magnetic field are shown in Fig. 7a and 7b, respectively [159 – 167]. Among all the materials reported so far, WTe$_2$, with its flexible properties, exhibits the highest Nernst Seebeck coefficients (7000 μV/K) [159]. Such high performance in this topological semimetal is primarily attributed to the high mobility of charge carriers within the system, favored by the sharp linear band dispersion. Charge carrier mobility and the Nernst signal are positively correlated. Therefore, a high Nernst signal is expected from semimetals with very low effective mass or massless fermions. Most of these materials exhibit high performance at very low temperatures, making them suitable for applications in cryogenic-based electronic devices.

A key advantage of Nernst-Ettingshausen effect-based devices is that they can be fabricated using a single material, whereas high-performance thermoelectric devices (shown in Fig 1b, and 1c) based on the longitudinal Seebeck effect (Fig. 1a) typically require both *p*-type and *n*-type materials [168, 169]. Therefore, these devices can be fabricated with simpler designs and smaller sizes, making them suitable for both macro- and nanodevices [168 – 169].

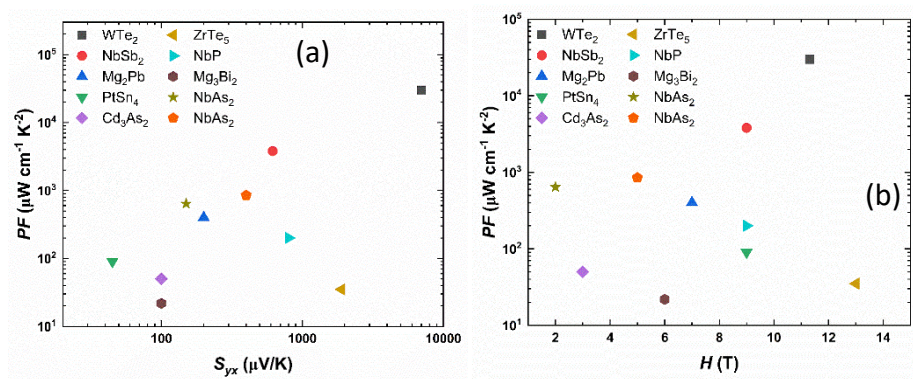

Fig. 7. (a) Nernst power factor as a function of Nernst Seebeck coefficient. (b) Nernst power factor as a function of applied magnetic field (B) for different Weyl semimetals.

To gain an idea about the effectiveness of the Nernst effect in thermal management applications over conventional thermoelectric materials based on the longitudinal Seebeck effect [60, 61, 67, 170 – 182], a comparative plot of power factor is shown in the figure. Among the several topological semimetals investigated, including $WTe_2$, NbP, $ZrTe_5$, $Cd_3As_2$, $NbSb_2$, and $NbAs_2$, $WTe_2$ exhibits the highest Nernst power factor [159 -167]. The Nernst power factor for $WTe_2$ is approximately 3 W/mK$^2$, which is typically 2-3 orders of magnitude higher than that of conventional semiconducting thermoelectric materials investigated in the past few decades.

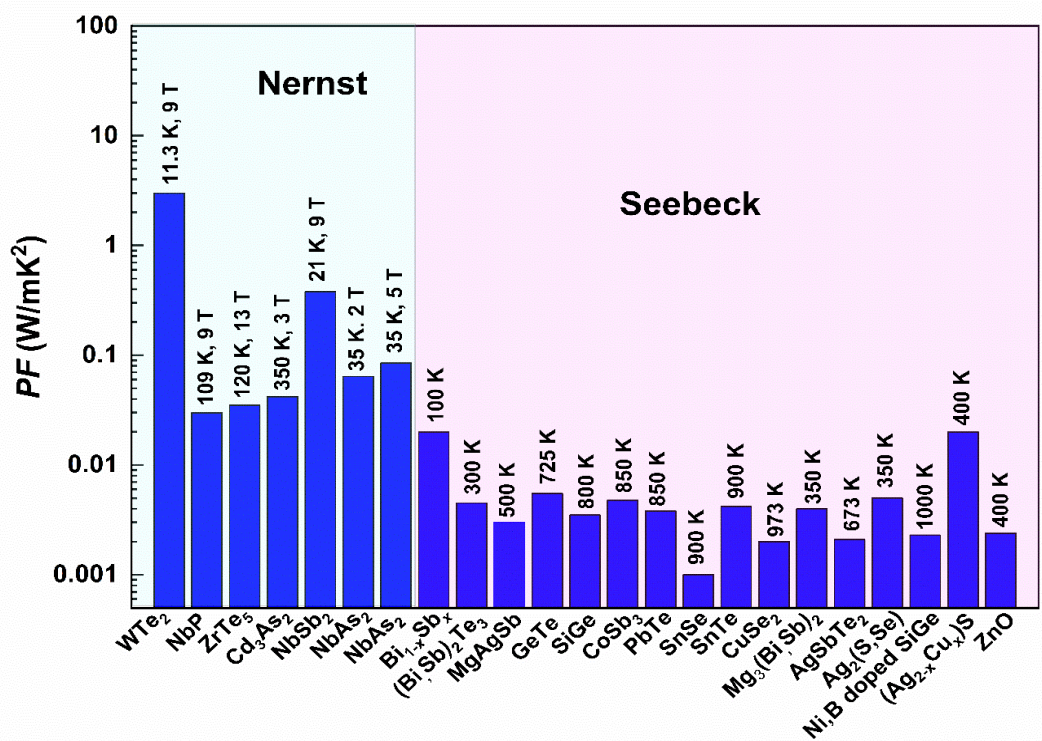

Fig. 8 Comparative plot of Nernst power factor for topological semimetals and Seebeck power factor of high-performance thermoelectric semiconductors.

For renewable energy generation and solid-state cooling applications, both power factor and figure-of-merit are important. For $WTe_2$, the measured thermal conductivity is reported as ~110 W/m·K [159] at a temperature of ~11 K, where the Nernst power factor reaches a maximum at a magnetic field of 9 T. The calculated figure of merit at 11.3 K is ~0.3, which is almost three times that of the *ZT* reported for $ZrTe_5$ (*ZT* = ~0.1 at 120 K) [164], but about two times lower than the values reported for $Cd_3As_2$ (*ZT* = ~0.7 at 350 K) [163]. Although the thermal conductivity of $WTe_2$ is significantly high, effective strategies such as alloying the main matrix, fabrication of thin films [183- 187] as they are helpful to tune the electronic and thermal properties of metals and semiconducting materials, can enhance phonon scattering, thereby reducing thermal conductivity and further improving the Nernst figure of merit.

As shown in Fig. 8, The second-best material showing the large Nernst power factor is $NbSb_2$. The reported PF for $NbSb_2$ is ~0.4 W/m.K$^2$ at temperature 21 K and magnetic field 9 Tesla [167]. The only material that has shown the Seebeck power factor in the same order of magnitude in the cryogenic temperature region is $Bi_{1-x}Sb_x$, without any applied external magnetic field, suggesting that for low-temperature (below 300 K) applications, topological semimetals are the most promising materials, whereas conventional semiconductors are suitable for temperatures above 300 K. This is mainly due to the fact that semiconductors with a narrow band gap contribute to electronic transport properties through thermally excited charge carriers, effectively, when the external temperature of the materials is raised to critical values.

Electronic band structure of the band gap semiconductor materials can be modified through the effective external dopant. Formation of the impurity peak in the density of states near valance/conduction band edge enhance the Seebeck coefficients leading to the large power factor. As a case study, highly insulating, both electrically and thermally, $Ag_2S$ materials, when doped with copper, $Ag_{2-x}Cu_xS$ [61], show an extraordinary improvement in the power factor, reaching ~0.2 W/m.K$^2$, which is the highest Seebeck power factor reported above room temperature from the band gap semiconductor. Among the inorganic materials, the ductile nature of $Ag_2S$ is highly beneficial for creating mechanically robust thermoelectric devices, including wearable thermoelectric generators for powering Internet of Things (*IoT*) applications [188], as well as solid-state cooling applications based on the Peltier effect. In addition to the Nernst effect, $WTe_2$ shows the multicomponent features such as an excellent ductile nature and very strong Ettingshausen effect. The Ettingshausen signal was measured to be about $5 \times 10^{-5}$ KA$^{-1}$m at a temperature of 23.1 K and a magnetic field of 9 Tesla. These findings suggest that $WTe_2$, with its exceptional Nernst-Ettingshausen performance and remarkable flexibility, holds great promise for the development of flexible and small-dimension thermoelectric devices.

## Conclusion

This chapter provides an overview of thermoelectric applications, focusing on waste heat recovery and solid-state cooling, with an emphasis on both the longitudinal Seebeck effect and the Nernst effect. The chapter begins with a general introduction to renewable energy and its significance for sustainable societal development. This is followed by a brief discussion of thermoelectric effects, including the Peltier effect, and an overview of state-of-the-art thermoelectric materials and their reported performance. The chapter then explores approaches for enhancing thermoelectric performance, including strategies for optimizing the figure of merit and conversion efficiency. The second half of the chapter elaborates into the fundamental aspects of topological quantum materials, specifically focusing on Weyl and Dirac semimetals. The basic working principles of the Nernst and Ettingshausen effects in these materials are discussed, along with an overview of state-of-the-art Weyl semimetal candidates. A comparison of the power factors achievable through the longitudinal Seebeck effect and the Nernst effect is presented to guide the selection of materials and device design strategies. Recent studies have shown that Weyl semimetals possess significant potential for thermal management applications. Therefore, further fundamental and applied research on these quantum material systems is crucial for advancing the development of efficient and sustainable thermoelectric technologies.